\documentclass{PoS}

\newcommand{\target}{{*}}

\newcommand{\msbar}{{\overline {\rm MS}}}
\newcommand{\csw}{{c_{\rm SW}}}
\title{2+1 flavor QCD simulation on a $96^4$ lattice}

\ShortTitle{2+1 flavor QCD simulation on a $96^4$ lattice}

\author{
 K.-I.~Ishikawa${}^{1,2}$,
 N.~Ishizuka${}^{3,4}$,
 Y.~Kuramashi${}^{2,3,4}$,
 Y.~Nakamura${}^{2}$,
 Y.~Namekawa${}^{4}$,
 Y.~Taniguchi${}^{3,4}$,
 \speaker{N.~Ukita}${}^{4}$,
 T.~Yamazaki${}^{2,3,4}$, and 
 T.~Yoshi\'e${}^{3,4}$\\
\hspace{5cm}(PACS Collaboration)\vspace{2mm}
\\
 ${}^1$Graduate School of Science, Hiroshima University, Higashi-Hiroshima, Hiroshima 739-8526, Japan\\ 
 ${}^2$RIKEN Advanced Institute for Computational Science, Kobe, Hyogo 650-0047, Japan\\
 ${}^3$Graduate School of Pure and Applied Sciences, University of Tsukuba, Tsukuba, Ibaraki 305-8571, Japan\\
 ${}^4$Center for Computational Sciences, University of Tsukuba, Tsukuba, Ibaraki 305-8577, Japan
E-mail: \email{ukita@ccs.tsukuba.ac.jp}}

\abstract{We generate $2+1$ flavor QCD configurations near the physical point on a $96^4$ lattice 
employing the 6-APE stout smeared Wilson clover action with a nonperturbative $c_{\rm SW}$ and the Iwasaki gauge action at $\beta=1.82$. 
The physical point is estimated based on the chiral perturbation theory using several data points generated by the reweighting technique from the simulation point, where $m_\pi$, $m_K$ and $m_\Omega$ are used as physical inputs. 
The physics results include the quark masses, the hadron spectrum, the pseudoscalar meson decay constants and nucleon sigma terms, using the nonperturbative
      renormalization factors evaluated with the Schr{\"o}dinger functional method.
}

\FullConference{The 33rd International Symposium on Lattice Field Theory\\
		14 -18 July 2015\\
		Kobe International Conference Center, Kobe, Japan}

\begin{document}

\section{Introduction}
\label{sec:intro}

Lattice QCD has today become the essential theoretical tool for quantitative studies in hadron physics. Once eliminating the systematic uncertainties including the extrapolation to the physical point, the simulation results are the predictions of QCD and can be compared with the experiment values. 
Thus, the recent lattice QCD simulations are performed near the physical point to control the  extrapolation to the physical point. 
In Refs. \cite{pacscs_chpt, pacscs_pp}, PACS-CS collaboration generated 
$2+1$ flavor QCD configurations in a lattice volume of spatial extent $L\approx 3$ fm at a lattice cutoff $a^{-1}\approx 2.2$ GeV, 
where the pion mass reached down to $m_{\pi}=145$ MeV. Moreover, the physical point simulation was realized by using the reweighting technique to adjust the hopping parameters to the physical point.     
Although the systematic uncertainty of the chiral extrapolation was much reduced in Ref. \cite{pacscs_chpt} and was fully eliminated in Ref. \cite{pacscs_pp},   
the spatial extent might be small, corresponding to $m_{\pi} L \approx 2$, a value for which finite volume effects will be difficult to be controlled.

In this work we generate $2+1$ flavor QCD configurations near the physical point on a larger lattice $(8\, {\rm fm})^4$ to reduce the finite volume effects and present the results by use of these configurations. 
The simulation is carried out with the 6-APE stout smeared Wilson clover action and the Iwasaki gauge action at a lattice spacing $a^{-1}\approx 2.3$ GeV by using K computer in RIKEN Advanced Institute for Computational Science. 
The pion mass in this simulation is 146 MeV and $m_{\pi}L\approx6$, which is expected to be large enough to ignore the finite volume effects within the statistical errors.   In order to obtain the results at the physical point, we generate new data points with the simulation point using the reweighting technique \cite{reweighting}. 
The pion mass for these reweighted points including the simulation point ranges from 144 MeV to 156 MeV. 
Using the ChPT formula to the chiral extrapolation for these data points, we determine the light quark masses and the lattice cutoff at the physical point, where we use $\pi, K$ and $\Omega$ masses as the physical inputs. 
The light hadron spectrum, the pseudoscalar decay constants and the nucleon sigma terms are also obtained.

This proceedings is organized as follows.
In Sec.~\ref{sec:detail} we present the simulation details of the configuration generation
and  of the data points generated by the reweighting technique from the simulation point.   
Section~\ref{sec:determination} is devoted to describe the determination of the physical point and the results. Our conclusions are summarized in Sec.~\ref{sec:conclusion}.

\section{Simulation details}
\label{sec:detail}

\subsection{Configuration generation}
\label{subsec:setups}

The use of smeared links in quark actions is currently an efficient way to perform simulations near the physical point because of better chiral symmetry and less exceptional configurations in the molecular dynamical steps than the use of thin links.  
We adopt the nonperturbatively $O(a)$-improved 2+1 flavor Wilson clover action with the 6-APE stout smeared links with the smearing parameter $\rho=0.1$ \cite{stout_smear} and the Iwasaki gauge action \cite{iwasaki}.

Simulations are performed on a $94^4$ lattice at $\beta=1.82$ which corresponds to the lattice cutoff $a^{-1}\approx 2.3$ GeV. We adopt a value of  the improvement coefficient $\csw=1.11$ which is determined  nonperturbatively in the Schr\"{o}dinger functional scheme. We choose  the hopping parameters $(\kappa_{\rm ud},\kappa_{\rm s})=(0.126\, 117, 0.124\, 790)$ to be near the physical point.     
The degenerated up-down quarks are simulated with the DDHMC algorithm \cite{luscher} using the even-odd preconditioning and the twofold mass preconditioning \cite{massprec1, massprec2}.  The strange quark is simulated with the UVPHMC algorithm \cite{fastMC,Frezzotti:1997ym,phmc,ishikawa_lat06} where the action is UV-filtered \cite{Alexandrou:1999ii} after the even-odd preconditioning without domain decomposition.

\begin{figure}[t!]
\begin{center}
\begin{tabular}{cc}
\includegraphics[width=70mm,angle=0]{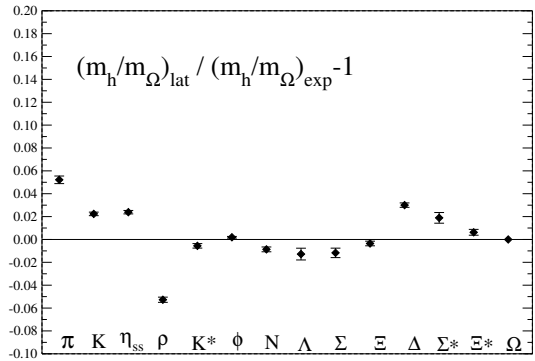}
\end{tabular}
\end{center}
\vspace{-.5cm}
\caption{Light hadron masses normalized by $m_{\Omega}$ at the simulation point in comparison with the experimental values.}
\label{fig1}
\end{figure}

After thermalization we generate 2000 trajectories and calculate hadronic observables solving quark propagators at every 10 trajectories. The statistical errors are estimated with the jackknife method with a binsize of 50 trajectories.  
Figure~\ref{fig1} compares the measured light hadron masses normalized by $m_{\Omega}$ with the experimental values. 
The results for $m_{\pi}/m_{\Omega}$ and $m_K/m_{\Omega}$ deviate from the experimental values: +5\% and +2\%, respectively. Thus, the simulation point is slightly away from the physical point.  
For other hadron masses we find less than 3\% deviation from the experimental values except for resonance states.

\subsection{Data points generated by reweighting technique}
\label{subsec:reweighting}

The reweighting factor for the up and down quarks is evaluated with a stochastic method introducing 
 a set of independent Gaussian random noises $\eta_i$ $(i=1,\dots,N_\eta)$ :
\begin{eqnarray}
 {\rm det}[W_{\rm ud}^2] =
 \left[\lim_{N_\eta\rightarrow \infty}
 \frac{1}{N_\eta} \sum_{i=1}^{N_\eta} 
  {\rm e}^{-\vert W_{\rm ud}^{-1}\eta_i\vert^2 +\vert \eta_i\vert^2}\right],
 \label{eq:rwfactor_ud}\quad
 W_{\rm ud} = \frac{D(\kappa_{\rm ud}^*)}{D(\kappa_{\rm ud})},
%\label{eq:w_ud}
\end{eqnarray}
where $D(\kappa_{\rm ud}^*)$ is the Wilson-Dirac matrix with a target hopping parameter $\kappa_{\rm ud}^*$. For the strange quark, further, we employ the square root trick, 
\begin{eqnarray}
 {\rm det}[W_{\rm s}] =
 \left[\lim_{N_\eta\rightarrow \infty}
 \frac{1}{N_\eta} \sum_{i=1}^{N_\eta} 
  {\rm e}^{-\vert W_{\rm s}^{-1}\eta_i\vert^2 +\vert \eta_i\vert^2}\right]^{\frac{1}{2}},
 \label{eq:rwfactor_s}\quad
 W_{\rm s} = \frac{D(\kappa_{\rm s}^*)}{D(\kappa_{\rm s})}.
%\label{eq:w_s}
\end{eqnarray}
To reduce the fluctuation in the stochastic evaluation (\ref{eq:rwfactor_ud}) and (\ref{eq:rwfactor_s}), we employ the determinant breakup technique \cite{pacscs_pp, hasenfratz, RBC_UKQCD}.

\begin{figure}[t!]
%\vspace{3mm}
\begin{center}
\begin{tabular}{cc}
\includegraphics[width=70mm,angle=0]{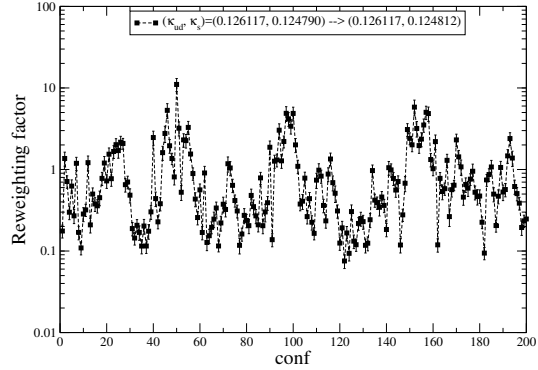}
\end{tabular}
\end{center}
\vspace{-.5cm}
\caption{Configuration dependence of reweighting factor from $(\kappa_{\rm ud},\kappa_{\rm s})=(0.126\,117,0.124\,790)$ to 
$(\kappa_{\rm ud}^\target,\kappa_{\rm s}^\target)=(0.126\,117,0.124\,812)$.}
\label{fig2}
\end{figure}

\begin{figure}[t!]
%\vspace{3mm}
\begin{center}
\begin{tabular}{ccc}
\includegraphics[width=50mm,angle=0]{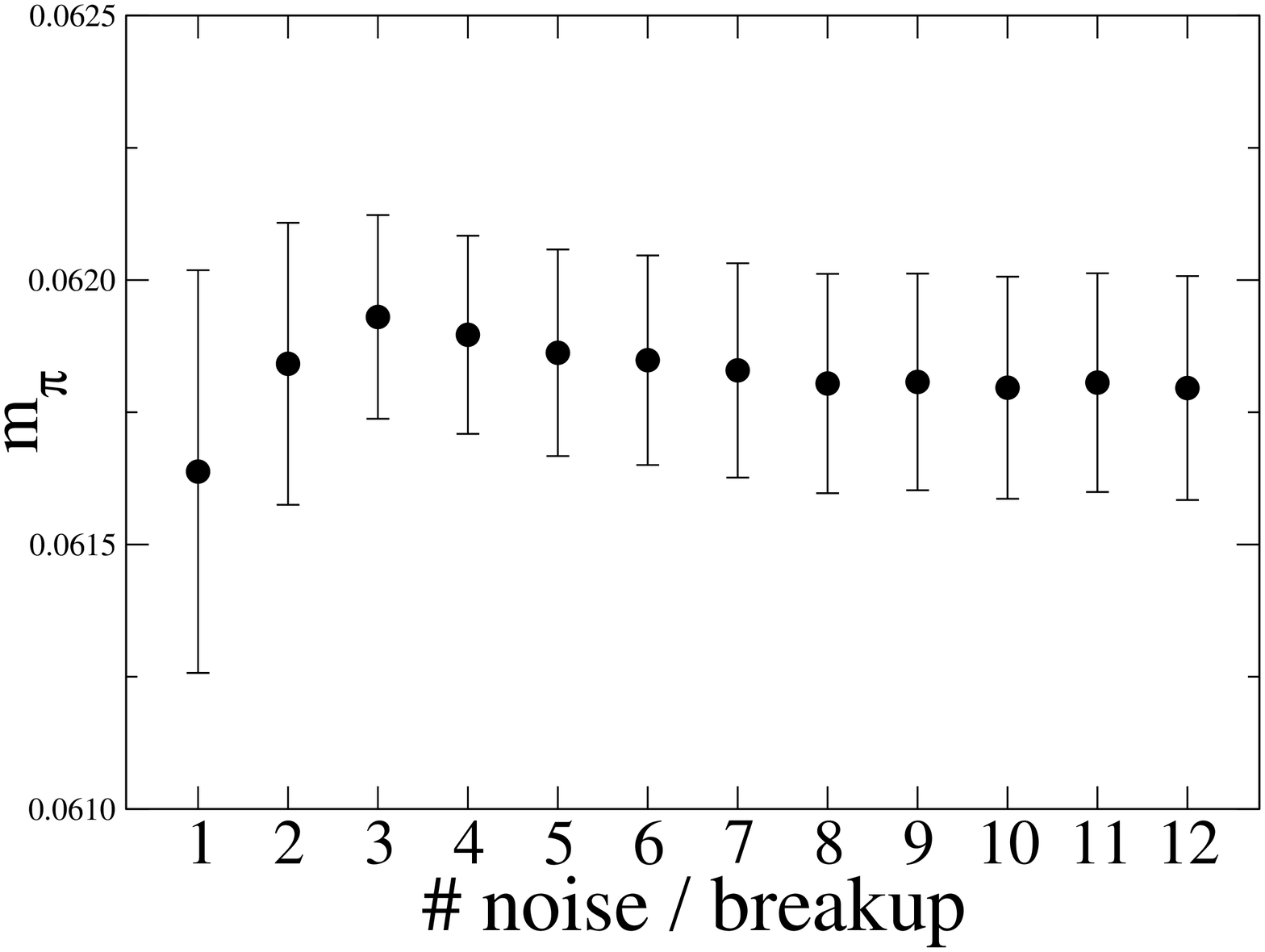}&
\includegraphics[width=50mm,angle=0]{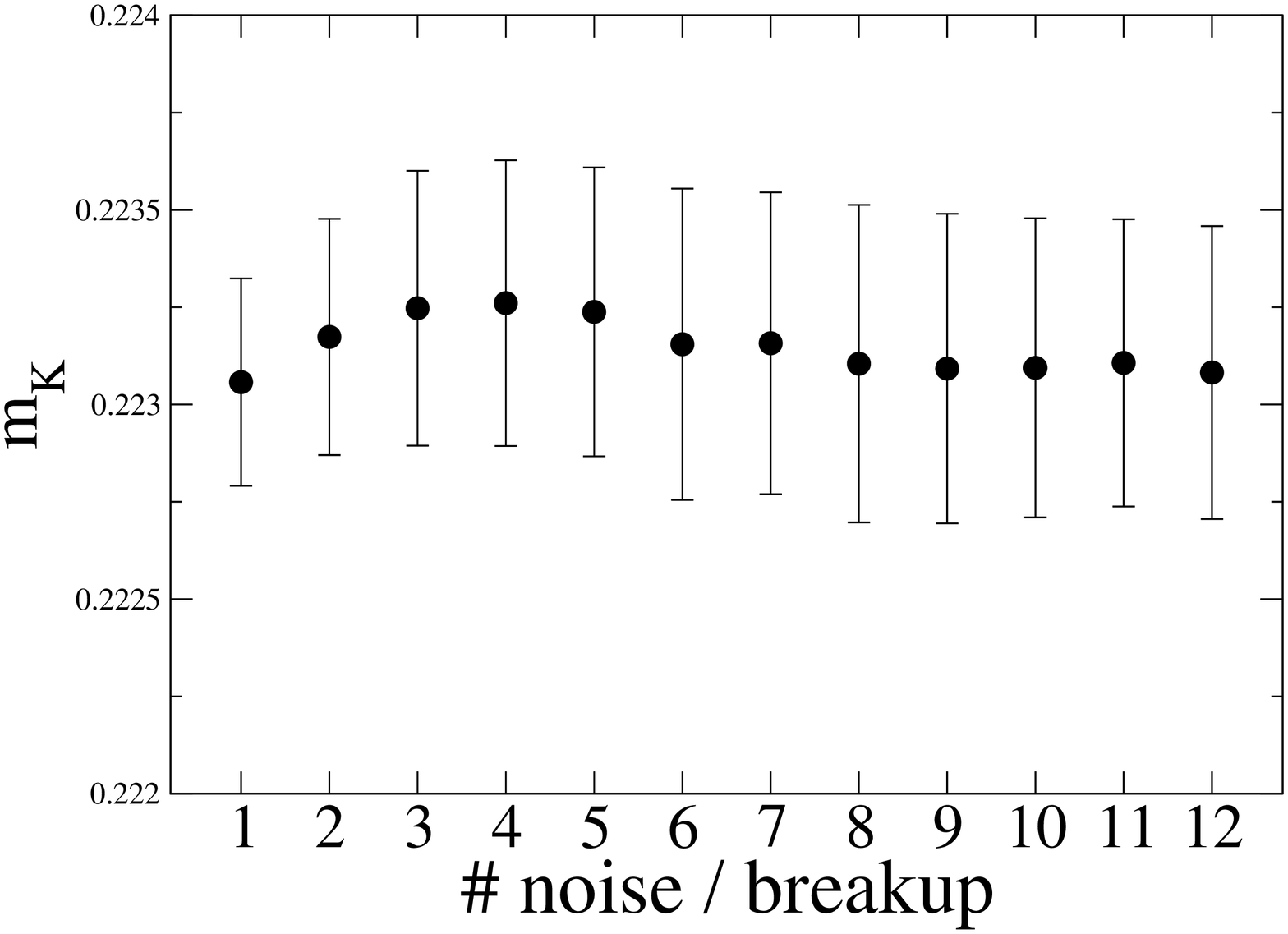}&
\includegraphics[width=50mm,angle=0]{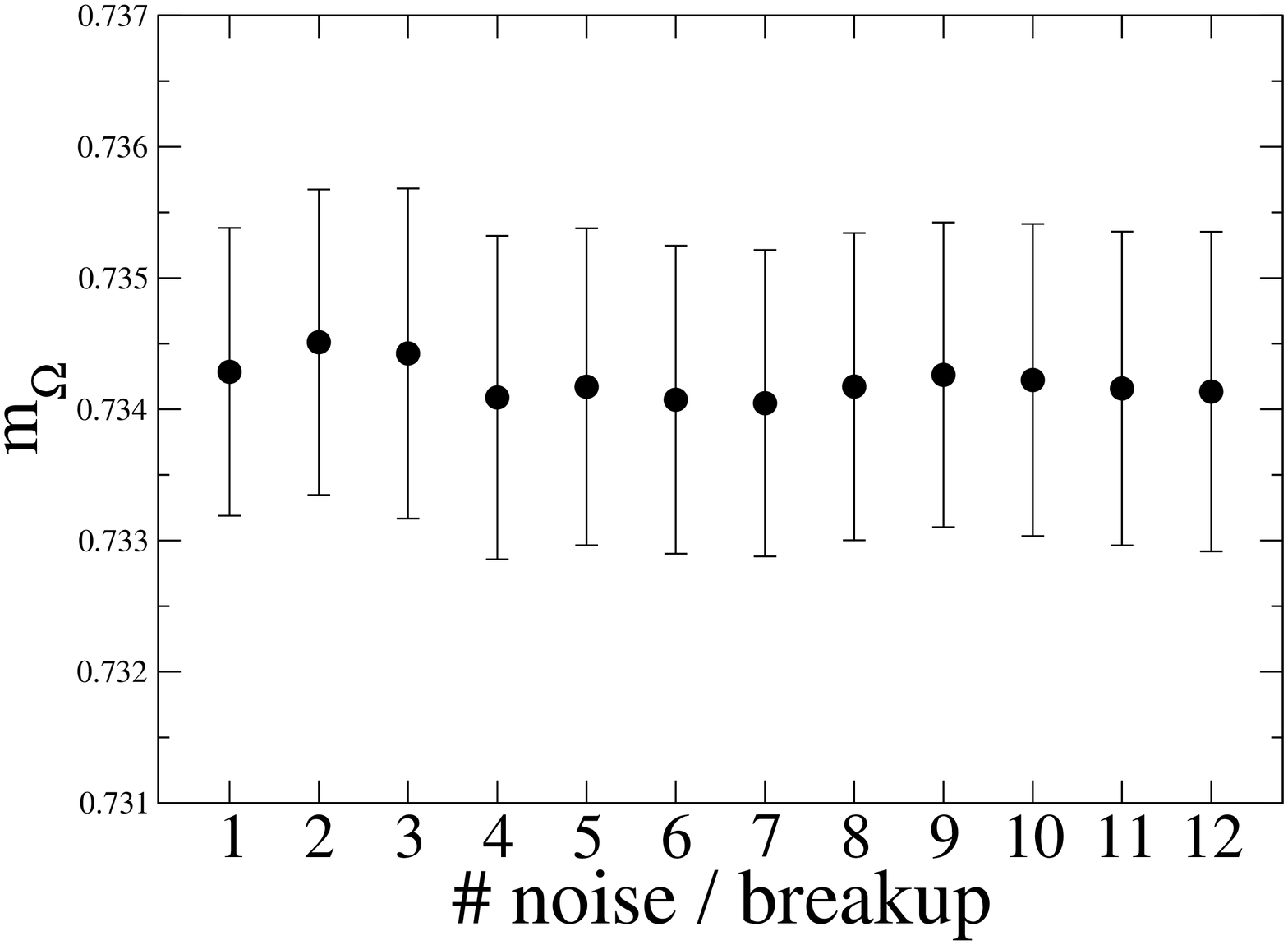}
\end{tabular}
\end{center}
\vspace{-.5cm}
\caption{$\pi$, $K$ and $\Omega$ masses at $(\kappa_{\rm ud}^\target,\kappa_{\rm s}^\target)=(0.126\,117,0.124\,812)$ as a function of the number of noises used in each determinant breakup.}
\label{fig3}
\end{figure}

We choose six target hopping parameters around the simulation point: $\kappa_{\rm s}=0.124\,768, 0.124\,812$ for $\kappa_{\rm ud}=0.126\,117$ and $\kappa_{\rm s}=0.124\,790, 0.124\,812, 0.124\,824, 0.124\,834$ for $\kappa_{\rm ud}=0.126\,111$.  We introduce 12 noises for each determinant breakup.  
Figure~\ref{fig2} shows the configuration dependence of the reweighting factor  from the simulation point $(\kappa_{\rm ud},\kappa_{\rm s})=(0.126\,117,0.124\,790)$ to 
$(\kappa_{\rm ud}^\target,\kappa_{\rm s}^\target)=(0.126\,117,0.124\,812)$, which is  normalized by the configuration average. The fluctuations are within a factor of two orders of magnitude. This is also the case for other reweighting factors.
Figure~\ref{fig3} shows $\pi$, $K$ and $\Omega$ masses at the reweighted point $(\kappa_{\rm ud}^\target,\kappa_{\rm s}^\target)=(0.126\,117,0.124\,812)$ as a function of the number of noises employed for the stochastic evaluation of each determinant breakup.
The results look converged from small number of noises.
This is the case for other hadron masses and for other reweighted points.
Therefore, 12 noises for each determinant breakup we used is enough to evaluate the reweighting factors. 
Here, we have the simulation point and six reweighted points to determine the physical point, where the pion mass ranges from 144 MeV to 156 MeV.

\section{Determination of the physical point and the results}
\label{sec:determination}

In order to obtain the results at the physical point, 
we determine the physical point
through 
the SU(2) ChPT formula for the degenerated up-down quark mass and make a linear extrapolation for the strange quark mass \cite{pacscs_chpt}. The formula is given by the following equations for $m_{\pi}^2/m_{\rm ud}$, $f_{\pi}$ and $f_K$ up to NLO:
\begin{eqnarray}
\frac{m_{\pi}^2}{2m_{\rm ud}} &=& 
 B\left\{1
 +\frac{m_{\rm ud}B}{8\pi^2f^2}\ln\left(\frac{2m_{\rm ud}B}{\mu^2}\right)
 +4\frac{m_{\rm ud}B}{f^2}l_3 \right\},\\
f_{\pi} &=& 
 f\left\{1
 -\frac{m_{\rm ud}B}{4\pi^2f^2}\ln\left(\frac{2m_{\rm ud}B}{\mu^2}\right)
 +2\frac{m_{\rm ud}B}{f^2}l_4 \right\}, \\
f_{K} &=& 
 \bar{f}\left\{1
 +\beta_f m_{\rm ud}
 -\frac{3m_{\rm ud}B}{32\pi^2f^2}\ln\left(\frac{2m_{\rm ud}B}{\mu^2}\right)\right\},
\end{eqnarray}
where $B=B^{(0)}_{\rm s}+m_{\rm s}B^{(1)}_{\rm s}$, $f=f^{(0)}_{\rm s}+m_{\rm s}f^{(1)}_{\rm s}$ and $\bar{f}=\bar{f}^{(0)}_{\rm s}+m_{\rm s}\bar{f}^{(1)}_{\rm s}$. 
There are nine unknown low energy constants in the expressions. 
For $K$ meson and $\Omega$ baryon, we use a simple linear formula with three unknown parameters,
\begin{eqnarray}
m_{K}^2 = \alpha_K + \beta_K m_{\rm ud} + \gamma_K m_{\rm s},\quad
m_{\Omega} = \alpha_{\Omega} + \beta_{\Omega} m_{\rm ud} + \gamma_{\Omega} m_{\rm s}.
\label{lin}
\end{eqnarray}
For other hadron masses, we adopt the same linear formula  with three unknown parameters. 
We need three physical inputs to determine the up-down and the strange quark masses and the lattice cutoff at the physical point. We choose $m_{\pi}$, $m_K$ and $m_\Omega$.

\begin{figure}[t!]
\begin{center}
\begin{tabular}{ccc}
\includegraphics[width=50mm,angle=0]{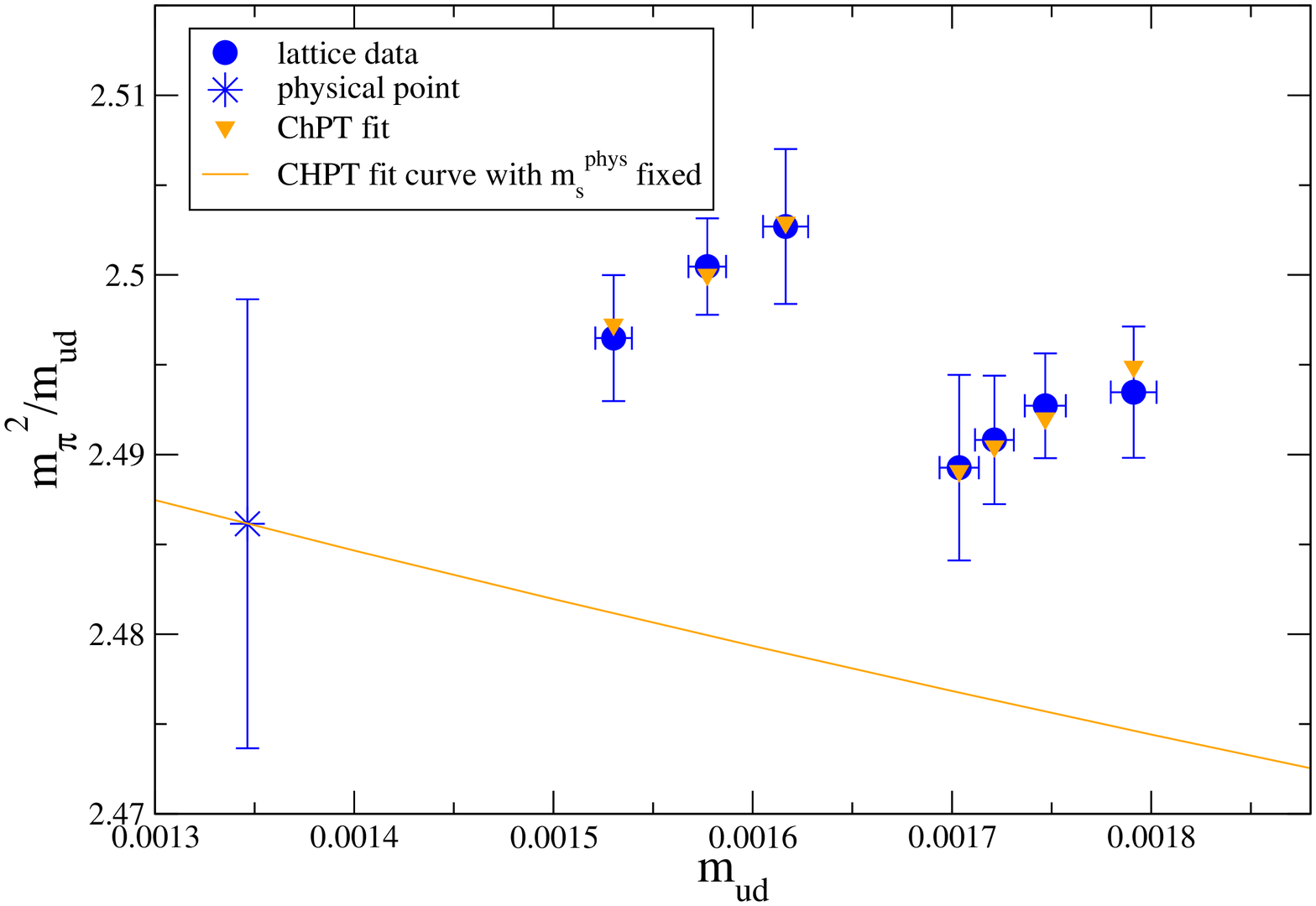}&
\includegraphics[width=50mm,angle=0]{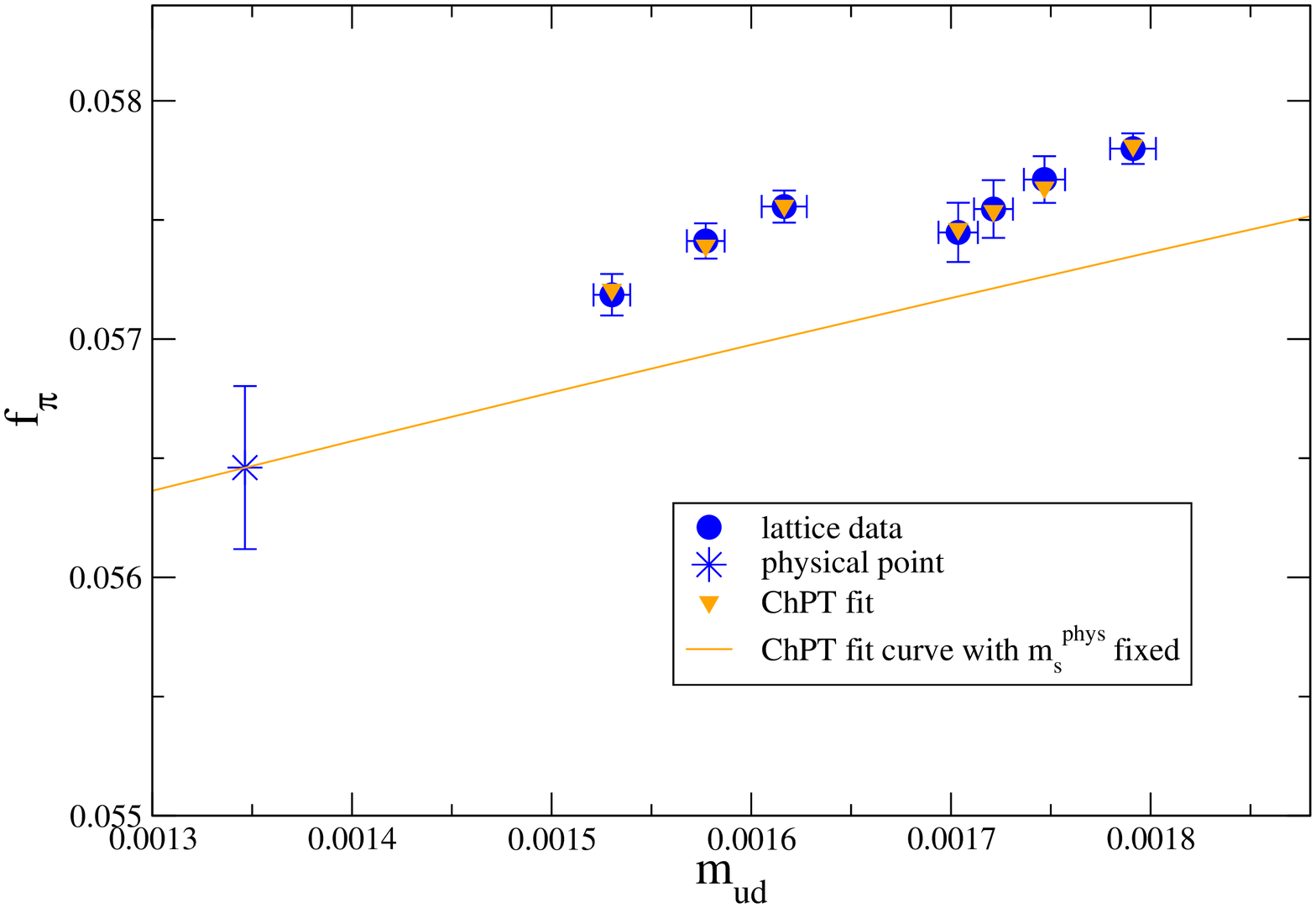}&
\includegraphics[width=50mm,angle=0]{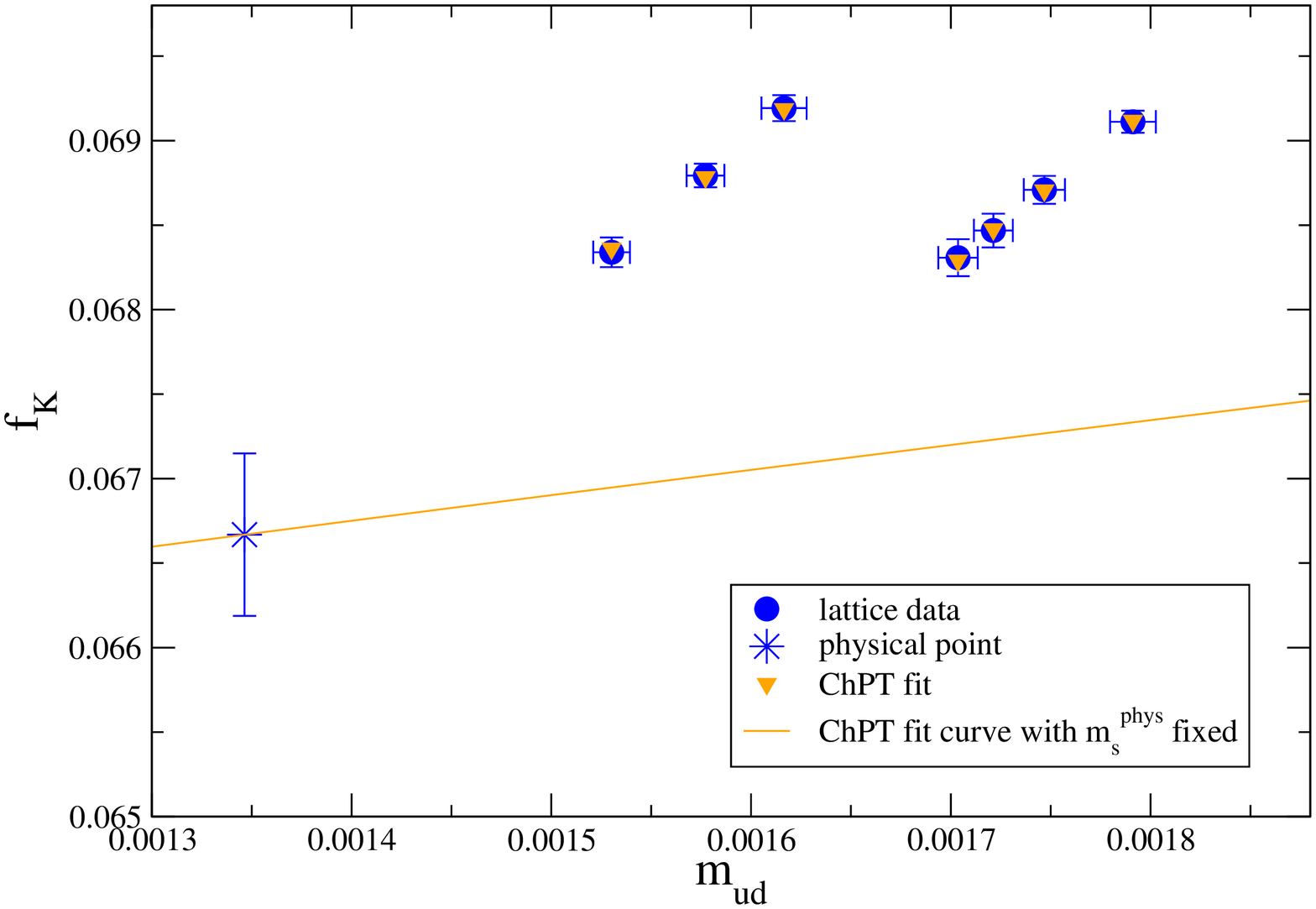}
\end{tabular}
\end{center}
\vspace{-.5cm}
\caption{SU(2) ChPT fit for $m_{\pi}^2/m_{ud}$ (left), $f_{\pi}$ (middle) and $f_{K}$ (right). Blue symbols represent lattice data. Orange triangles represent the ChPt fit. Orange lines  represent the ChPT fit curves with the physical strange quark mass fixed.}
\label{fig4}
\end{figure}

Figure~\ref{fig4} shows the quark mass dependence of ${m_{\pi}^2}/{m_{\rm ud}}$, $f_{\pi}$ and $f_K$.   
They show that the quark mass dependence are reasonably described by the ChPT formula. 
We find that the low energy constants at NLO defined at the renormalization scale $m_{\pi}$, that is   $\bar{l}_3$ and $\bar{l}_4$,  are determined with acceptable errors,
\begin{eqnarray}
 \bar{l}_3=2.87(62),\quad \bar{l}_4=4.38(33).
\end{eqnarray} 
Our result is consistent with $N_{f}=2+1$ average in FLAG2013~\cite{flag}, $\bar{l}_3=3.05(99)$ and $\bar{l}_4=4.02(28)$. 

After the additional linear fits for $K$ and $\Omega$, we obtain at the physical point
\begin{eqnarray}
m^{\overline{\rm MS}}_{\rm ud}=3.142(26)(35)(28) {\rm MeV},\ \  
m^{\overline{\rm MS}}_{\rm s}= 88.59(61)(98)(79) {\rm MeV},\ \ 
a^{-1}= 2.333(18) {\rm GeV}, 
\end{eqnarray}
where the first errors in the quark masses are the statistical ones and  the second and the third errors are coming from the renormalization factor $Z_m^\msbar=0.9950(111)(89)$  at $\mu=2$ GeV nonperturbatively determined in the Schr{\"o}dinger functional scheme~\cite{z}.

For the pseudoscalar decay constants, we obtain 
\begin{eqnarray}
f_{\pi} = 131.79(80)(90)(1.25) {\rm MeV},\ \  
f_{K} = 155.55(68)(1.06)(1.48) {\rm MeV},
\end{eqnarray}
 with the second and the third errors coming from the renormalization factor $Z_A=0.9650(68)(95)$ which is  nonperturbative  determined in the Schr{\"o}dinger functional scheme~\cite{z}. These values are consistent with experiment~\cite{pdg}. 
We also obtain the ratio $f_K/f_{\pi}=1.1808(50)$ and observe 3.1$\sigma$ deviation including the systematic error from the experiment~\cite{pdg}.

\begin{figure}[t!]
\begin{center}
\begin{tabular}{cc}
\includegraphics[width=70mm,angle=0]{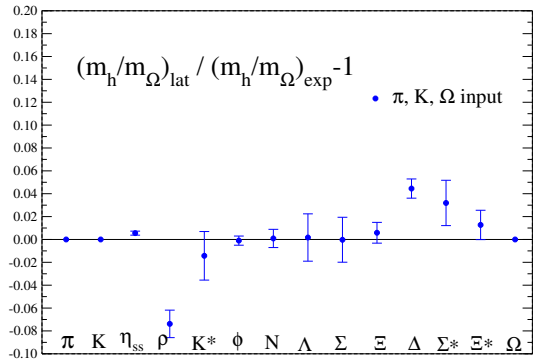}
\end{tabular}
\end{center}
\vspace{-.5cm}
\caption{Hadron masses normalized by $m_{\Omega}$ at the physical point in comparison with experiment.}
\label{fig5}
\end{figure}

Figure~\ref{fig5} shows the light hadron spectrum extrapolated to the physical point normalized by $m_{\Omega}$ in comparison with experiment. 
We find less than 5\% deviation from the experimental values. 
Stable particles in QCD, which are the octet baryons are consistent with the experimental  values within the errors. 
For unstable particles in QCD ($\rho, \Delta,\cdots$), we observe deviations from the experimental values. We need further investigations of those as resonances \cite{resonance, cppacs_rho}.

\begin{figure}[t!]
\begin{center}
\begin{tabular}{cc}
\includegraphics[width=70mm,angle=0]{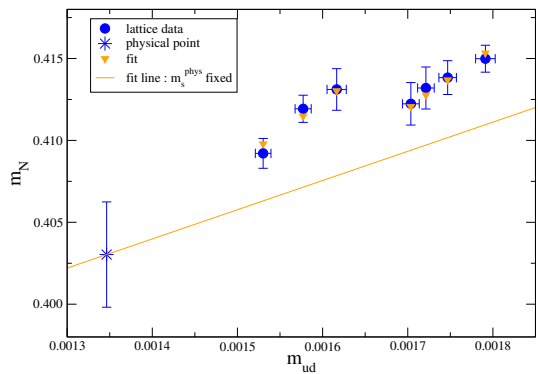}
\end{tabular}
\end{center}
\vspace{-.5cm}
%\caption{Linear chiral extrapolation for the nucleon mass. Symbols represent  the same as Figure~\ref{fig4}.}
\caption{Linear chiral extrapolation for the nucleon mass. Symbols represent  the same as Figure~4.}
\label{fig6}
\end{figure}

Finally, we show the results for nucleon sigma terms. 
For the nucleon mass denoted by blue symbols in Figure~\ref{fig6}, we assume a simple linear formula with three unknown parameter as eq. (\ref{lin}). 
We obtain 
\begin{eqnarray}
\sigma_{\rm ud}=55(13) {\rm MeV},\ \  
\sigma_{\rm s}=107(73) {\rm MeV},
\end{eqnarray}
through the Feynman--Hellmann theorem.


\section{Conclusion}
\label{sec:conclusion}
We have generated $2+1$ flavor QCD configurations near the physical point on a $96^4$ lattice 
employing the 6-APE stout smeared Wilson clover action with a nonperturbative $c_{\rm SW}$ and the Iwasaki gauge action at $\beta=1.82$. 
The physical point is estimated based on the SU(2) ChPT formula using several data points generated by the reweighting technique from the simulation point. Adopting $m_{\pi}$, $m_K$ and $m_{\Omega}$ as the physical inputs, we obtained 
the physics results including the quark masses, the hadron spectrum, the pseudoscalar meson decay constants and nucleon sigma terms.

\begin{acknowledgments}
The configuration generation has been carried out
on the K computer  at RIKEN Advanced Institute for Computational Science. 
The hadron measurements have been carried out on the HA-PACS and the T2K-Tsukuba at Center for Computational Sciences, University of Tsukuba. 
This work is supported in part by MEXT SPIRE Field 5, JICFuS (hp120281, hp130023, hp140209) and Grants-in-Aid for Scientific Research
from the Ministry of Education, Culture, Sports, Science and Technology
(Nos. 24740143, 15K05068).
\end{acknowledgments}

\end{document}